\documentclass{aastex}
\usepackage{spr-astr-addons}
\usepackage{amsmath}
\newcommand{\rthis}[1]{\textcolor{black}{#1}}
\usepackage{multirow}
\usepackage{url}
\usepackage[plainpages=false, colorlinks=true, anchorcolor=blue, linkcolor=blue, citecolor=blue, bookmarks=false]{hyperref}

\RequirePackage{color}
\begin{document}

\title{Two Dimensional Clustering  of Gamma-Ray Bursts using durations and hardness}
\slugcomment{Not to appear in Nonlearned J., 45.}
%% Running heads
\shorttitle{Classification of GRBs}
\shortauthors{A. Bhave, et al}
\author{Aishwarya Bhave\altaffilmark{1}}
\and \author{Soham Kulkarni\altaffilmark{2}} \and  \author{Shantanu Desai\altaffilmark{3}}
\and  \author{P.K. Srijith \altaffilmark{4}}
\altaffiltext{1}{Department of Information Technology, NIT
Raipur, Chhatisgarh-492010, India}
\altaffiltext{2}{Department of Physics, University of Florida, Gainsville, FL 32611, USA}
\altaffiltext{3}{Department of Physics, IIT Hyderabad, Kandi, Telangana-502285, India}
\altaffiltext{4}{Department of Computer Science and Engineering, IIT Hyderabad, Kandi, Telangana-502285, India}

%\email{ep13b1003@iith.ac.in}

\begin{abstract}
Gamma-Ray Bursts (GRBs) have been traditionally  divided   into two categories:  ``short''  and ``long'' with durations less than and greater than two seconds, respectively.  However, there is a lot of literature (with conflicting results) regarding the existence of a third intermediate class. To investigate this issue, we carry out a   two-dimensional classification using  the  GRB hardness and duration, and also incorporating the uncertainties in both the variables, by using an extension of Gaussian Mixture Model called Extreme Deconvolution (XDGMM).
We carry out this  analysis on datasets from two detectors, viz. BATSE and Fermi-GBM. We consider the duration and hardness features in log-scale for each of these datasets and determine the best-fit parameters using XDGMM. This is  followed by information theoretic criterion-based tests (AIC and BIC) to determine the optimum number of classes. For BATSE, we find that both AIC and BIC show preference for two components with close to  decisive and  decisive significance, respectively. For Fermi-GBM, AIC shows preference for three components with decisive significance, whereas BIC does not find any significant difference between two and three components. Our analysis codes have been made publicly available.

\end{abstract}

\keywords {GRB Classification; Bayesian Information Criterion; Akaike Information Criterion}

\section{Introduction}

Gamma-ray bursts (GRBs)  are short-duration energetic cosmic explosions with prompt emission  between keV-GeV energies,  which  are being continuously detected at the rate of about one per day ~\citep{Kumar15,Zhang16,Schady}. They are located at cosmological distances, although a distinct signature of cosmological time dilation in the GRB light curves remains elusive~\citep{Singh}. The first convincing case for bifurcating the GRB population into two categories was made from an  analysis of the BATSE data~\citep{Kouveliotou}, and led to
establishing the  conventional classification of GRBs into short
(T90 $<$ 2 s) and long (T90 $>$ 2 s) classes, where T90 is the time which encompasses 90\% of  the burst's fluence, and is usually used as a proxy for  the duration of a GRB. Note however that this boundary between short and long GRBs is known to be detector dependent~\citep{Bromberg,Tatro15}.
Short GRBs are usually associated with  binary neutron star  (or other compact object) mergers~\citep{Nakar} and long GRBs with core collapse supernovae~\citep{Woosley}. However, there are known exceptions to the above general picture~\citep{Zhang,Perna18,Amati20,Ahumada}.
%Most classification studies of GRBs have been done using T90, although other measures have also been  proposed~\citep{Zhang06,Li16}.  

Despite the conventional wisdom of  only two distinct  GRB classes,  multiple groups have argued  over the years for the existence of an intermediate class of GRBs in between the short and long bursts, using T90 as the criterion for classification.
The first such claim for an intermediate-duration GRB class, with T90 in the range between 2 and 10 seconds in the BATSE dataset was  put forward by ~\citet{Horvath98}  and \citet{Mukherjee}, and  subsequently confirmed by the analysis of the complete BATSE    dataset~~\citep{Horvath2002,Chattopadhyay,Horvath19}. 
%This  picture is also muddled by evidence for core collapse supernova been associated with  a short GRB (GRB 200826A)~\cite{Amati20}.

However, this has been  disputed by ~\citet{Zitouni}, who  found  that two lognormal distributions fit the BATSE T90 data  much better compared to three components.  Evidence for a third lognormal component was also found in the BAT data of Neil Gehrels SWIFT Observatory (Swift, hereafter) ~\citep{Horvath2008, Zhang08, Huja09, Horvath10,Horvath16,Zitouni,Tarnopolski,VonKienlin}. However, these results have been disputed by other authors, who found that the T90 distribution prefers two components~\citep{Yang,Tarnopolski19}.  ~\citet{Kulkarni17} carried out a unified classification of the T90 distributions for  the GRB datasets from BATSE, Fermi, Swift, and Beppo-Sax, and found that among these, only for Swift GRBs in the observed frame is the evidence for three classes marginally significant at about $2.4\sigma$. However,  when the same analysis is done for the Swift GRBs in the intrinsic GRB frame, two components are preferred. For all other datasets, evidence for three components is either very marginal or disfavored. Some other works have pointed out that the third component could be an artifact of the skewness in the distribution of long GRBs. If the data  is modelled using skewed distributions, only two distributions are sufficient~\citep{Tarnopolski16b,Kwong,Tarnopolski19,Tarnosai}.

Extension of studies on GRB classification  using both duration and hardness,  as well as in dimensions greater than two  have 
also not reached a common consensus.~\citet{Horvath06} and ~\citet{Chattopadhyay} argued for three components in the BATSE GRB data using two-dimensional clustering in T90-hardness and T90-fluence planes respectively. ~\citet{Chattopadhyay17,Maitra17} have argued for five components  in the BATSE data by using multiple clustering techniques on six different variables. For Swift data,~\citet{Veres} showed using multiple clustering techniques   that three components are favored  in the two dimensional  $\log$(T90) -$\log$ (hardness) plane, and the intermediate class has overlap with X-ray flashes. 
%~\citep{Horvath2008, Huja09, Horvath10} using duration and also from two-dimensional clustering using both duration and hardness~\citep{Veres}.
However, these results are in conflict with more recent analysis by ~\citet{Jiang}, who showed by applying two dimensional Gaussian Mixture Models (GMM) on T90 and hardness ratio on Swift GRBs, that the data favor only two components instead of three. Most recently, ~\citet{Tarno21} showed by applying  graph theory techniques  to  GRB hardness and duration, that the data is consistent with two groups, although a third group cannot be ruled out. Using linear discriminant analysis,~\citet{Zhang22} showed that a linear combination of the duration, fluence, and peak flux is a better discriminant between long and short bursts for Fermi-GBM data. They also showed that long GRBs could also be further sub-divided into long-bright GRBs and long-faint GRBs~\citep{Zhang22}. A recent summary of all results on GRB classification can be found in ~\citet{Tarnobatse19}. Most recently, Swift GRBs have also been classified into categories based entirely on prompt light curves~\citep{Jespersen}.

To resolve this imbroglio, we use two-dimensional clustering in the hardness vs T90 plane to  find out the optimum number of GRB classes. One difference compared to all previous works on GRB classification is that we incorporate the uncertainties in the GRB hardness and T90, while doing the classification. We apply an extension of the GMM,  which incorporates the uncertainties in the data~\citep{Bovy}. We then uniformly apply this method to  the latest available GRB data from BATSE and Fermi-GBM detectors,  based on the observed  T90 and hardness along with their associated uncertainties.

For model comparison, we use two widely used information theoretic criteria, viz. Akaike Information Criterion (AIC)  and Bayesian Information Criterion (BIC). 
Both of these information-criterion based model comparison techniques have been applied to a variety of problems in astrophysics and particle physics, including in the classification of GRBs~\citep{Shi,Desai16a,Desai16b,Kulkarni17,Ganguly17,Kulkarniexoplanet,KrishakOJA,KrishakDAMA,Krishakanais,Krishak} and references therein. 
%However, as  one deficiency in all the aforementioned works on GRB classification, is that the uncertainties in the duration as well as the hardness have not been taken into account. In this work, we carry out a two-dimensional classification using GRB duration and hardness and by incorporating the uncertainties.

The outline of this paper is as follows. The datasets  used for our analysis are described in Sect~\ref{sec:1}. The analysis methodology and model comparison techniques are outlined in Sect.~\ref{sec:analysis}. We then present our results for the various GRB datasets in Sect.~\ref{sec:results}, including a very brief comparison with previous results.  We conclude in Sect.~\ref{sec:conclusions}.

%\section*{}
%\label{sec:intro}

%\footnote{http://gammaray.msfc.nasa.gov/batse/grb/catalog/current} 4B catalog~\citep{BATSE} and Fermi-GBM\footnote{http://heasarc.gsfc.nasa.gov/W3Browse/fermi/fermigbrst}

\section{Datasets}
\label{sec:1}

Herein, we consider the GRB datasets available from BATSE~\citep{BATSE} and Fermi-GBM~\citep{GBM,fermi2020}.   Among these, Fermi detector is still online and detecting on the order of about  one new GRB per day.  We did not consider other catalogs such as those from Beppo-Sax and INTEGRAL,  since they either contained  a very small sample of GRBs or did not have any publicly available data for the hardness of the observed bursts. We did not consider the bursts from RHESSI, as it has no onboard GRB triggering mechanism. Although there have been many previous works which used the Swift GRB catalog for classification~\citep{Horvath2008,Horvath16,Yang,Kulkarni17}, we note that  Swift is much more sensitive in the softer bands, meaning that the short GRBs are under-abundant in the sample, which affects the fits~\citep{Band06}. The selection effects for Swift are therefore difficult to account for~\citep{Coward}. Therefore, we do not use the data from  Swift for our analysis. Before describing each of these datasets, we first explain how the  hardness variable was estimated.

\subsection{Hardness Definition}
\label{sec:hardness}
Spectral hardness (also referred to as hardness ratio) ($H_{obs}$, hereafter)  of a GRB is  defined as the ratio between the GRB fluences in different energy bands. For BATSE, we use $H_{obs}$ as the ratio between the $50-100$ keV and the $20-50$ keV bands. The BATSE catalog provides errors for the fluence in both these bands. From this, one can estimate the error in  $\log (H_{obs})$  from error propagation. 
%For SWIFT we used the ratio between  the $50-100$ keV and the $25-50$ keV bands. 
For Fermi-GBM, we consider the ratio of the fluences in 50-300 keV and 10-50 keV to calculate the hardness. We have obtained the errors in hardness and T90 from the latest Fermi-GBM  catalog (Narayana-Bhat, private communication, 2020).

Note that many previous works have used BATSE fluences in the 100-300 and 50-100 keV bands for the hardness ratio~\citep{Mukherjee,Horvath06,Chattopadhyay,Tarnobatse19,Horvath19}.  Here, instead we used the fluences in the 20-50 keV energy range in the denominator of the hardness ratio, in order to have overlap in a similar energy range as Fermi-GBM (10-50 keV). Although, we are not analyzing Swift data in this work, clustering studies based on hardness with Swift have also used a similar energy band (25-50 keV)~\citep{Yang}. Therefore, choosing the BATSE fluence in the 20-50 keV range easily allows us to compare with the results from other detectors and also complement the above studies which have used the fluences in 100-300 keV energy range.

\subsection{BATSE dataset}
The current BATSE  GRB~\citep{BATSE} catalog contains 2041 GRBs detected between 1991 and 2000 with duration information and a total of 2035 events with flux and fluence information. Among these, 1973 GRBs contained both duration and fluence information, of which we omitted 39 GRBs for which the fluences in either the 20-50 keV or in the 50-100 keV were less than or equal to zero. Therefore, we are left with 1934 GRBs for our classification purposes.  The average fractional error in T90 is 18\%, whereas the average fractional error for the fluence in the 50-100 keV and 20-50 keV energy bands  is equal to 12\% and 17\%, respectively.The average and median fractional error in the hardness ratio is 36\% and 10\%, respectively.

\subsection{Fermi-GBM dataset}
As of  June 2020, Fermi-GBM released their  fourth catalog containing 2356 GRBs~\citep{Gruber,VonKienlin,GBM,fermi2020}. Among these, T90, hardness ratio (as defined above), along with their associated uncertainties are available for 2330 GRBs.
Of these, we used 2329 GRBs for analysis, since one GRB had negative value for the hardness and hence could not be used for the analysis. The hardness ratio was calculated using the ratio of the background subtracted flux spectrum (using photon counts) in 50-300 keV to that in 10-50 keV, and averaging over the detectors. More details on this catalog can be found in ~\citet{fermi2020}. The average fractional error in T90 and hardness ratio is equal  to 10\% and 18\%, respectively. \rthis{The median fractional error in the hardness ratio is equal to 3\%.}

\section{Methodology}
\label{sec:analysis}
Extreme Deconvolution (XDGMM, hereafter) is an extension of GMM~\citep{Feigelson17}, which takes into account the uncertainty in the observed data~\citep{Bovy,astroML,Wechsler}. It has been used for a variety of applications in astrophysics such as velocity distribution from Hipparcos data~\citep{Bovy}, classification of pulsars~\citep{Reddy},   the three-dimensional motions  of the stars in Sagittarius streams ~\citep{sagittarius_koposov},  classification of  neutron star masses~\citep{Keitel}, detection of dark matter subhalo candidates~\citep{Miguel}. We provide a brief description of the XDGMM method using the same notation as ~\citet{Reddy}.

We assume that the noisy dataset $x_i$ is related to the true values $v_i$ as follows~\citep{Bovy,astroML}:
\begin{equation}
x_i=R_i v_i + \epsilon_i,
\label{eq:transform}
\end{equation}
where $R_i$ is the rotation matrix used to transform the correct values to the observed noisy dataset.   Similar to ordinary GMM, we assume that the probability density  of the true values $v$ can be written as a mixture of $K$ Gaussians given by
\begin{equation}
p(v_i) = \sum_{j=1}^K \alpha_j \mathcal{N} (v_i|\mu_j, \Sigma_j)   
\end{equation}
where $\mu_j$ and $\Sigma_j$ denote the means and variances of each of the Gaussian distribution, whereas $\alpha_j$ denotes the weight of each Gaussian subject to $\sum \limits_{j=1}^K \alpha_j=1$.
Note that $x_i$
and $v_i$ could be multi-dimensional vectors, and for our example, denote the 2-D dataset comprising of the natural logarithms of  T90 and $H_{obs}$.
We consider the noise $\epsilon_i$ (in Eq.~\ref{eq:transform}) to be a Gaussian random variable with zero mean and variance equal to  $S_i$.  The likelihood of the model parameters ($\theta \equiv$ \{$\alpha$, $\mu$, $\Sigma$, $R_i$, $S_i$\}) for each noisy data point ($x_i$) can then be written as~\citep{Bovy}:
\begin{equation}
p(x_i|\theta)=\sum_{j=1}^{K} \alpha_j \mathcal{N}(x_i|R_i\mu_j,R_i\Sigma_jR_i^{T}+S_i)
\end{equation}

The last step in XDGMM is to maximize the likelihood of the dataset with respect to the model parameters. This can be done (as in GMM) by adding   the individual log-likelihood functions:
\begin{equation}
\underset{\theta}{\operatorname{argmax}} \;   L  = \sum_{i=1}^{N} \ln (p(x_i|\theta)),
\end{equation}
where $N$ is the total number of datapoints.
This objective function  is maximized using an extension of the \rthis{Expectation-maximization} algorithm~\citep{Bovy}.  Similar to GMM, XDGMM returns a likelihood, which can then
be utilized for model comparison.

\subsection{Fitting method}
\label{sec:param}
We now apply XDGMM to the GRB dataset using the $\log$ (T90) and $\log H_{obs}$  as inputs, where $\log$ refers to natural log. We use the XDGMM implementation in the {\tt astroML} module~\citep{astroML}.  We stack the $\log$ (T90) and $\log (H_{obs})$. Their uncertainties constitute the diagonal elements of their respective covariance matrices, with non-diagonal elements kept at zero, since the errors between different GRBs are independent. The stacked covariance matrices are fed to the XDGMM  algorithm, whose output consists of 
the weights, means, and covariances for the input number of clusters

XDGMM by itself does not determine the optimum number of GRBs, which is an input parameter to the algorithm. For finding the optimum number of components, we  apply XDGMM by varying the  number of GRB components,
 and then use model selection techniques, as discussed in the next section to determine the optimum number of clusters.

The comparison of models based on the difference in likelihood after finding the best-fit parameters for each model is not the optimum way to find the correct number of components, even though this has been used in the GRB classification literature~\citep{Horvath06,Horvath10}.
For  mixture models, minus twice log of difference in likelihood does not asymptote to the usual $\chi^2$   distribution~\citep{Ghosh85,Feigelson17}. Only when the variances are equal does this statistic follow the $\chi^2$ distribution~\citep{Feigelson17}.

Furthermore, even though the value of the likelihood increases, the addition of extra  free parameters leads to increased model complexity  and is generally undesired. 
Therefore, the additional free parameters need to be  penalized or taken into account so as to avoid overfitting.  To address these issues,
a number of both frequentist and Bayesian model-comparison techniques have been used
over the past decade to determine the best model which fits the observational data~\citep{Liddle,Liddle07,Lyons,Weller,Krishak}. Here, we use information criteria based tests such as  AIC and  BIC  for  model comparison, since these are straightforward to compute from the likelihoods (which are returned as one of the outputs from XDGMM).
AIC/BIC have also been previously used  for GRB classification by a number of authors~\citep{Mukherjee,Tarnopolski16b,Tarnopolski,Jiang,Kulkarni17,Tarnopolski19,Zhang22}. More information about AIC and BIC and its application to astrophysical problems can be found in ~\citet{Liddle,Liddle07,Sharma,Krishak}.

\subsection{AIC}

The AIC is used for model comparison, when we need to penalize for any additional free parameters  to avoid overfitting. A preferred model in this test is the one with the smaller value of AIC between the two hypothesis. The AIC is given by,

\begin{equation}
AIC = 2p - 2 \ln L .
\label{eq:aic}
\end{equation}

\noindent where $p$ is the number of free parameters in the model and $L$ is the likelihood.
The second term favors models with high value of likelihood, while the first term penalizes models which uses large number of parameters. Models with large number of parameters might have a high likelihood but will over fit on the data.    The absolute value of AIC is usually not of interest. The goodness of fit between two hypothesis (A) and (B) is described by the difference of the AIC values and is given by,

\begin{equation}
\Delta AIC = AIC_{A} - AIC_{B},
\end{equation}
\noindent where  $AIC_{A}$ - $AIC_{B}$ correspond to the AIC values for the  hypothesis A and B. \citet{Burnham} have provided qualitative
strength of evidence rules to assess the significance of a model based on the $\Delta$AIC  values between the two models. $\Delta$ AIC$<2$ corresponds to substantial support, those with $4 \le \Delta$AIC$ \le 7$ have considerably less support, and those with  $\Delta$AIC have virtually no support. Therefore, 
 $\Delta$AIC$>10$ is considered as decisive evidence against the model with higher AIC~\citep{Liddle07,Krishak}. 
  
% Should we just define model a to be the 2G model and B as the 3G model? Because in the table 1 description, the values and the description(caption) above contradict. I've highlighted that in red.\\
\footnote{To avoid any ambiguity in our representation of our results, we have consistently kept the 3-Gaussian model as the null hypothesis, which simplifies the analysis and makes a positive value of $\Delta AIC$,  favor the 3-Gaussian and a negative value favors the 2-Gaussian.}

\subsection{BIC}

The BIC is also used for penalizing the use of extra parameters. As in the case of AIC, the model with the smaller value of BIC is the preferred model. The penalty in the BIC test is harsher than that in the case of AIC and is given by,

\begin{equation}
BIC = p \ln N - 2 \ln L .
\label{eq:BIC}
\end{equation} 

The first  term in Eq.~\ref{eq:BIC}  acts as a very harsh measure needed for the BIC test. The goodness of fit used
for hypothesis testing between two models $A$ and $B$ is given by,
\begin{equation}
\Delta BIC = BIC_{A} - BIC_{B} .
\end{equation}
Similar to AIC, the model with lower value of BIC is favored. To assess the significance of a model, strength of evidence rules have also been proposed based on $\Delta$BIC using Jeffreys scale~\citep{Robert,Liddle07}. $\Delta$ BIC $> 5$ is considered as strong evidence and $\Delta$ BIC $> 10$ is considered as decisive evidence in favor of the model with the smaller BIC value.

%We note that AIC and BIC are two complementary criteria for model comparison~\citep{Liddle,Liddle07,Tarnopolski16b,Burnham}. The penalization term in BIC is greater than AIC for $N>8$. Therefore, the penalization is more stringent in BIC for large samples. AIC tries to select a model that most adequately describes reality. On the contrary, BIC tries to find the true model among the set of candidates. BIC  has a tendency to underfit, while AIC, as a
%more liberal method, is inclined towards overfitting. Therefore, sometimes they lead to contradictory results as  they try to satisfy different conditions. More details on the difference between AIC and BIC can be found in ~\citet{Liddle,Liddle07,Tarnopolski16b,Burnham,Krishak}.

\section{Results}
\label{sec:results}
We apply the techniques discussed in the previous sections to the GRB datasets from  BATSE and Fermi-GBM. We find the mean value of  $\log$ (T90) and $\log (H_{obs})$ and its standard deviation for each GRB class, by varying the total number of components from one to five, followed by  maximizing the likelihood using XDGMM for each of  the hypothesis. This choice for the number of GRB components is large enough, as it  allows to easily discern the minimum value of AIC/BIC and also allows us to cross-check the results of some works, which have found upto 5 GRB components~\citep{Chattopadhyay17,Maitra17}.   Using these best-fit parameters, we then implement the information criterion based model-comparison techniques to determine the optimum number of components

%The GMM and the corresponding parameter estimation using the EM algorithm are implemented using the {\tt sklearn.mixture} module of the  python library {\tt Scikit-learn}. Covariance types {\tt full} and {\tt tied} are used for generating the model with the number of components ranging from one to five and we choose the covariance type, which yields the maximum value of likelihood. Covariance type {\tt full} corresponds to the case when each GMM component has its own covariance matrix, whereas the {\tt tied} corresponds to the case when all components share the same general covariance matrix.

%For all the detectors except Fermi, the maximum value is achieved for covariance type equal to {\tt full}. Note that our main goal is to try to ascertain whether a three-component fit is favored compared to a two-component one or vice-versa. Therefore, we are agnostic to the  value for  the number of components for which  we get  the minimum value of AIC or BIC, although we do report its value for all the detectors.

%examples of how to add figures.
\subsection{BATSE}
\iffalse
The current BATSE  GRB~\citep{BATSE} catalog contains 2041 GRBs detected between 1991 and 2000 with duration information and a total of 2035 events with flux and fluence information. Among these, 1973 GRBs contained both duration and fluence information, of which we omitted 39 GRBs for which the fluences in either the 20-50 keV or in the 50-100 keV were less than or equal to zero. Therefore, we are left with 1934 GRBs for our classification purposes.  The average fractional error in T90 is 18\%, whereas the average fractional error for the fluence in the 50-100 keV and 20-50 keV energy bands  is equal to 12\% and 17\%, respectively.The average and median fractional error in the hardness ratio is 36\% and 10\%, respectively.
\fi
%We then apply the parameter estimation procedure using XDGMM  outlined in Sect.~\ref{sec:param}. 

\begin{figure}
\includegraphics[width=8cm]{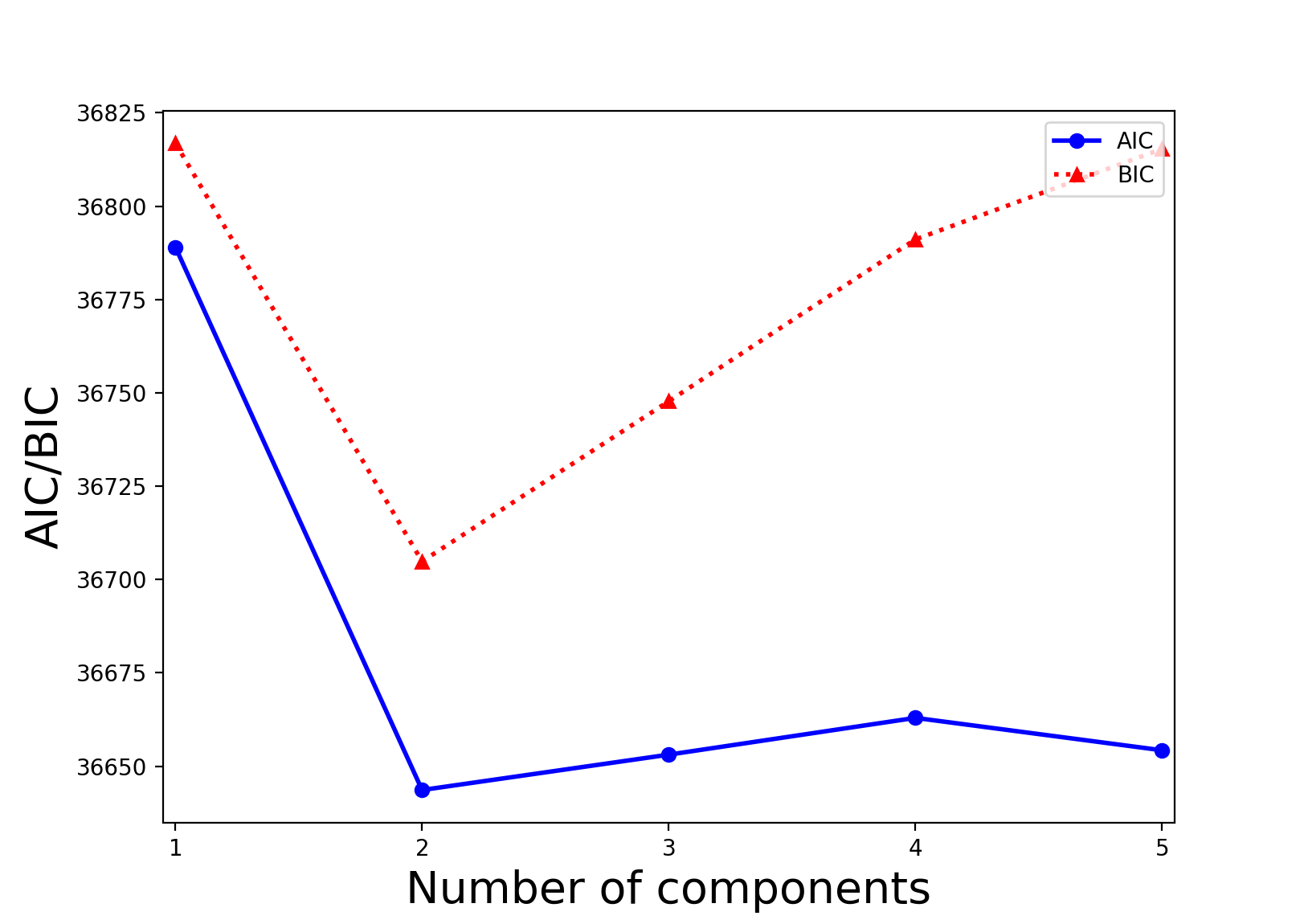}
\caption{AIC and BIC values  as a function of the number of  Gaussian components for BATSE data after two-dimensional clustering. Both AIC and BIC favor two components with near decisive or decisive significance, respectively.}
\label{fig:batse2}
\end{figure}

\begin{figure}
\includegraphics[width=8cm]{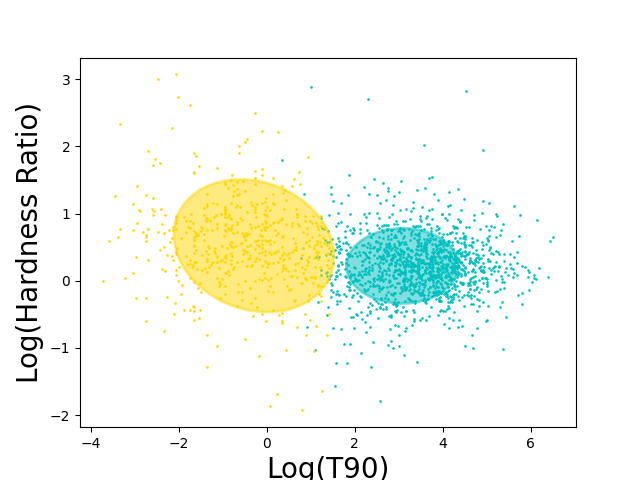}
\caption{Scatter plot of $\log H_{obs}$ vs $\log$ (T90) (expressed in seconds) for BATSE data. The ellipses indicate the 1$\sigma$ contours for two components, using our XDGMM based analysis and are centered on the best-fit parameters obtained from Table~\ref{tab:batse}.}
\label{fig:batse1}
\end{figure}

\begin{figure}
\includegraphics[width=8cm]{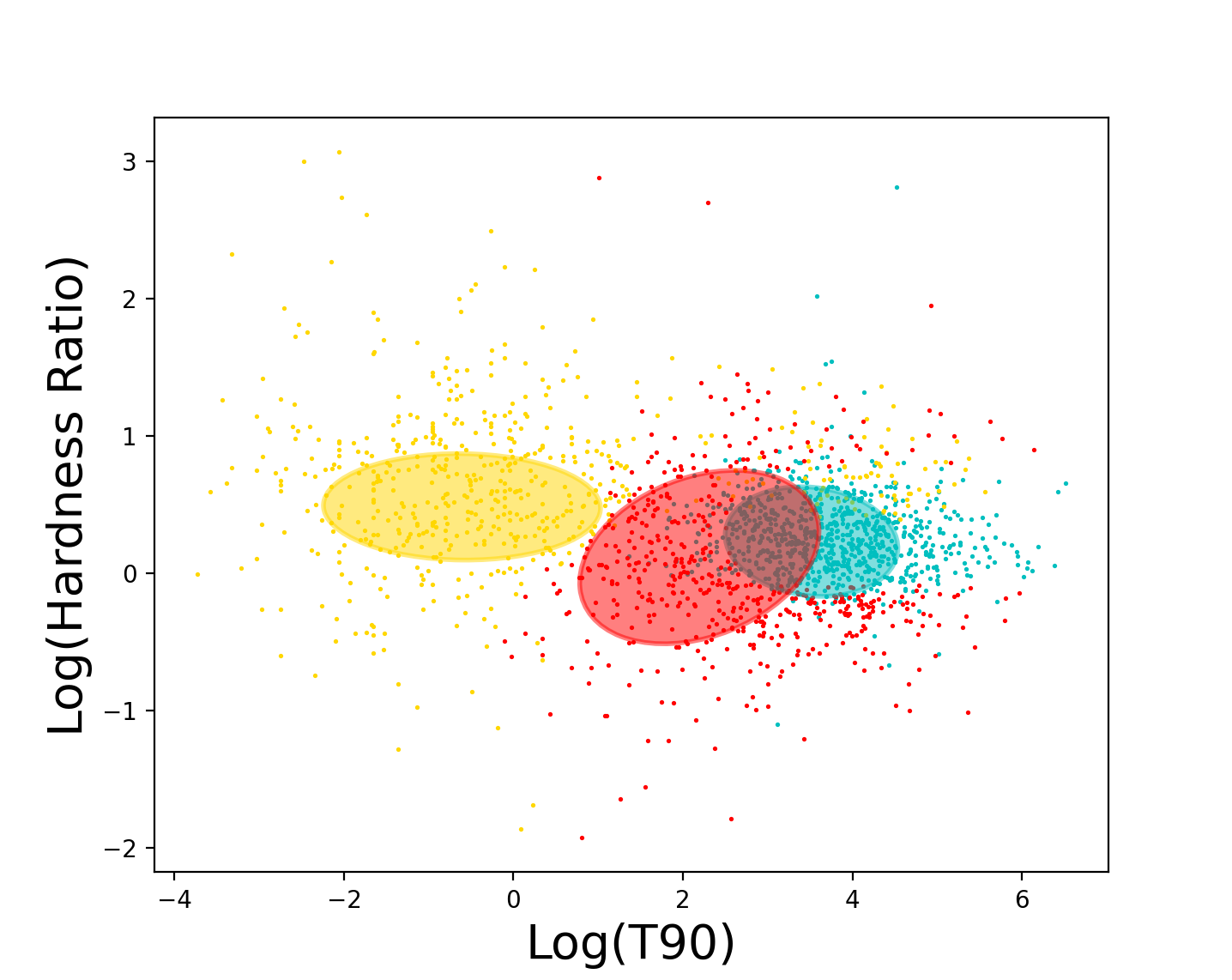}
\caption{Scatter plot of $\log (H_{obs})$ ratio vs $\log$ (T90) (expressed in seconds) for BATSE data. The ellipses indicate the 1$\sigma$ contours from our XDGMM based analysis assuming three components and  centered on the best-fit parameters obtained from Table~\ref{tab:batse}.}
\label{fig:batse3}
\end{figure}

A complete summary of the results on applications of XDGMM to the BATSE GRB dataset, including the best-fit parameters and their covariance matrices are shown in Table~\ref{tab:batse}. While fitting for two components, we find that 808 and 1126 GRBs belong to the short and long category respectively.  When we fit for three components we
find a total of 689, 762, and 483 GRBs in the short, intermediate, and long categories respectively.
The AIC and BIC plots as a function of the number of components can be found in Fig.~\ref{fig:batse2}. Here, both AIC and BIC prefer two components. The $\Delta$BIC value crosses the threshold of 10, needed for decisive evidence. The $\Delta$AIC value is also close to 10. Therefore, both AIC and BIC results using XDGMM are broadly in agreement and favor two GRB categories.
The $1\sigma$ ellipses for two  and three components can be found in Fig.~\ref{fig:batse1} and Fig.~\ref{fig:batse3}, respectively.   %The means for the 2-component model are given by [[ 3.08  0.22], [-0.29  0.52]] (cf. Table~\ref{tab:batse}) For the 3-component model, the means are given by [[2.12 0.2] [-0.54 0.61] [3.4 0.21]], where the first two elements refer to natural log of T90 and $H_{obs}$.

There is more than 20 years of literature on classification of BATSE GRBs. However, the results from different works are not in accord with each other, with  disparate statistical techniques often leading to opposite conclusions.  
So, we only compare our results to  a few selected  papers,  where both T90 and hardness (or other fluence related parameters) are used for classification. Results of classification of BATSE GRBs using only T90 are summarized in ~\citet{Kulkarni17}. The first cogent case for three GRB classes in BATSE data using spectral information, was made by ~\citet{Misra}, who used two multivariate clustering methods using $K$-means partitioning and Dirichlet mixture modeling using fluence vs T90 to argue for three components. However, no estimate of the significance was made.
Around the same time, an analysis similar to this using GMM in the log (T90)- log( $H_{obs}$) plane was done by ~\citet{Horvath06}, who found that three components were favored using frequentist model comparison by evaluating chi-square probability with the addition of the third component. They also found an anti-correlation between the duration and the hardness.  
% ~\citet{Chattopadhyay17} usedGMM based analysis on six different variables (two durations, peak flux, total fluence, and two spectral hardness ra- tios), five types of bursts are preferred. However,~\citet{Horvath19} showed that the five groups found in ~\citet{Chattopadhyay17} are not new classes of GRBs, but a consequence of the sub-classification of the short and intermediate groups using the peak flux in 256-ms timescale. This was also confirmed using fuzzy clustering techniques~\citep{Modak}. 
Most recently, ~\citet{Tarnobatse19} caried out multi-variate modelling of BATSE data using symmetric as well as  skewed distributions.  He showed that the T90 distribution is consistent with two components. In two dimensional space, 
almost all the distributions are consistent with two components, except when flux was used for classification in which case, the distribution was well-fitted by three-component Gaussian or $t$-distributions. A comprehensive summary of their classification results in higher-dimensional spaces  can be found in Table 1 of ~\citet{Tarnobatse19}. This work argued that it is not possible to unequivocally prove the existence of a third GRB component, since the third component   could be a  spurious artifact caused by  the finite size of the sample and due  to a particular realization of the random
sample that could bias the results.

The results from  our analysis, which incorporates the errors in T90 and $H_{obs}$  support two classes.

\begin{table*}[!htbp]
\caption{Results from model comparison for BATSE GRBs. The first column contains the total number of GRB classes.
The two component array ($\log$ T90, $\log H_{obs}$) in the  second column denotes the best-fit values for the mean value of the logarithm of  T90 (in seconds) and the  logarithm  of  $H_{obs}$. The $2 \times 2$   matrix in the third column indicates the covariance matrix $\Sigma$ returned by XDGMM. The fourth column ($n_i$) denotes the  total number of GRBs. These columns have been shown separately for two and three classes of GRBs. The next set of columns show the AIC and BIC values for each GRB category.   The last two columns indicate the $\Delta$AIC, and $\Delta$BIC between the three component and two-component model, which are used for model comparison.   In this table, the preferred value for every test is highlighted in bold. We note that $\Delta$AIC = AIC (2 components) - AIC (3 components) and same for $\Delta$BIC. Therefore, if  $\Delta$AIC or $\Delta$BIC$>0$, then two GRB classes are preferred and vice-versa. We note that AIC and BIC prefer two components with near decisive and decisive  significance, respectively.}
\label{tab:batse}
\begin{tabular}{|c|ccc|cc|cc|}
\hline
$k$ & $\mu$ & $\Sigma$ & $n_{i}$ &   AIC  &  BIC   & $\Delta(AIC) $& $ \Delta(BIC)$ \\
\hline
\multirow{2}{*}{2} & (3.08,0.22) &  $\left(\begin{array}{cc}                            0.8078 &0.0026\\0.0026 & 0.1565 \\               \end{array}\right)$ & 1126       & \multirow{2}{*}{\textbf{36643.6}} & \multirow{2}{*}{\textbf{36704.9}} & \multirow{5}{*}{-9.5} & \multirow{5}{*}{-42.9} \\
\cline{2-4}
& (-0.29,0.52) &$\left(\begin{array}{cc}      1.6557 & -0.1494\\ -0.1494& 0.4862 \\     \end{array}\right)$ & 808 & &  & &\\
\cline{1-6}
\multirow{3}{*}{3} &(2.12,0.20) &$\left(\begin{array}{cc}       1.0358 &0.0936\\ 0.0936 & 0.3560 \\     \end{array}\right)$ & 762     & \multirow{3}{*}{36653.1} & \multirow{3}{*}{36747.7}  & & \\
\cline{2-4}
& (-0.54,0.61) &$\left(\begin{array}{cc}       1.3979 &-0.0453\\ -0.0453& 0.3925 \\     \end{array}\right)$ & 689 & &  & & \\
\cline{2-4}
& (3.40,0.21) &$\left(\begin{array}{cc}       0.5599 &-0.0021\\ -0.0021& 0.1134 \\     \end{array}\right)$& 483 & &  & & \\
\hline
\end{tabular}
%\tablecomments{In the table the preferred value for every test is highlighted in bold.}
\end{table*}

\subsection{Fermi-GBM}
\iffalse
As of  June 2020, Fermi-GBM released their  fourth catalog containing 2356 GRBs~\citep{Gruber,VonKienlin,GBM,fermi2020}. Among these, T90, hardness ratio, along with their associated uncertainties are available for 2330 GRBs.
Of these, we used 2329 GRBs for analysis, since one GRB had negative value for the hardness and hence could not be used for the analysis. The hardness ratio was calculated using the ratio of the background subtracted flux spectrum (using photon counts) in 50-300 keV to that in 10-50 keV, and averaging over the detectors. More details on this catalog can be found in ~\citet{fermi2020}. The average fractional error in T90 and hardness ratio is equal  to 10\% and 18\%, respectively.
\fi

\begin{figure}
\includegraphics[width=8cm]{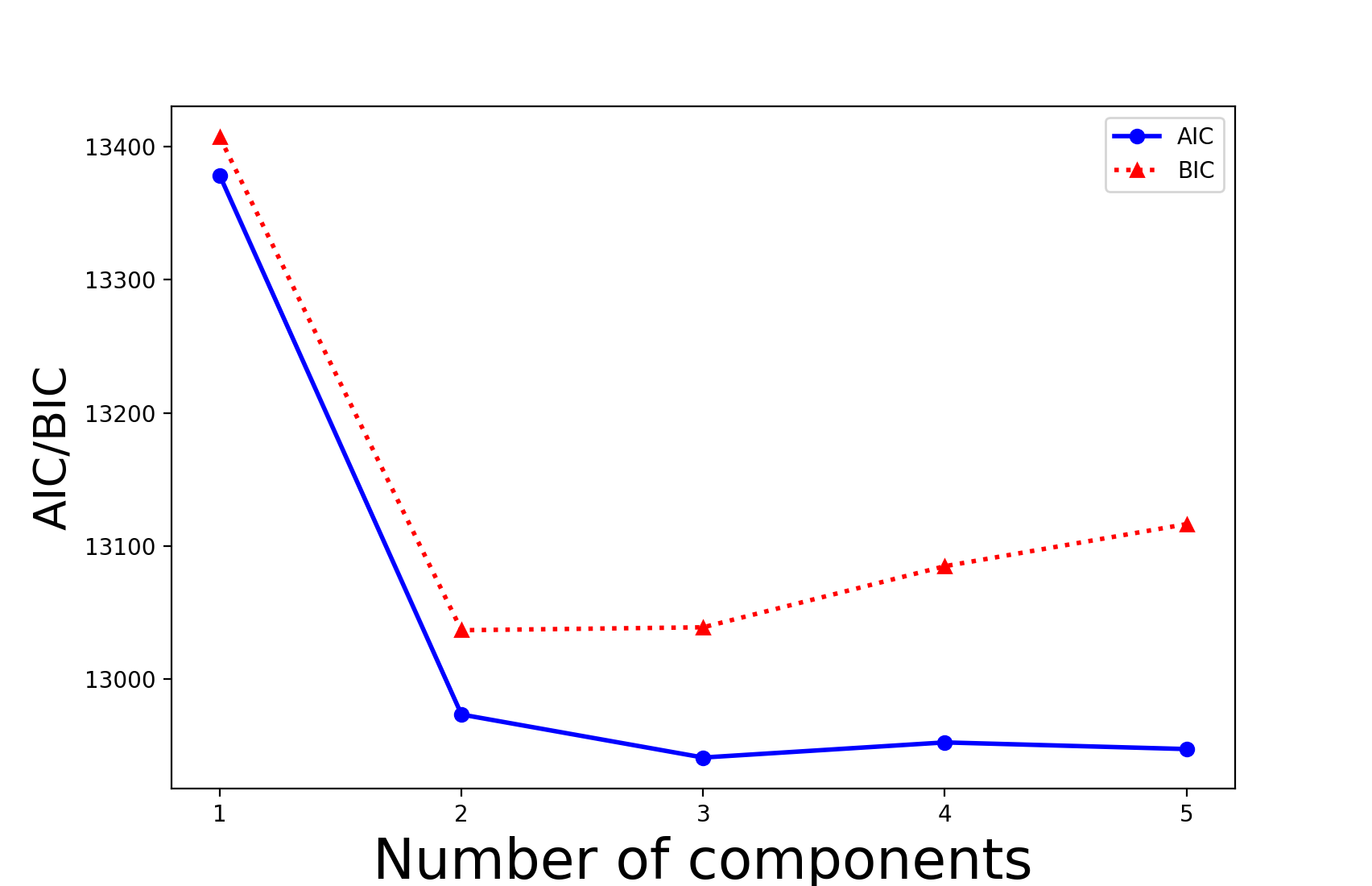}
\caption{AIC and BIC values  as a function of the number of  Gaussian components for Fermi-GBM data after two-dimensional clustering. The minimum value of BIC is obtained for two components. However the difference compared to the same for three components is negligible. For AIC, the minimum value is obtained for three components and the difference with respect to two components points to decisive significance in favor of the three component model.}
\label{fig:fermi2}
\end{figure}

\begin{figure}
\includegraphics[width=8cm]{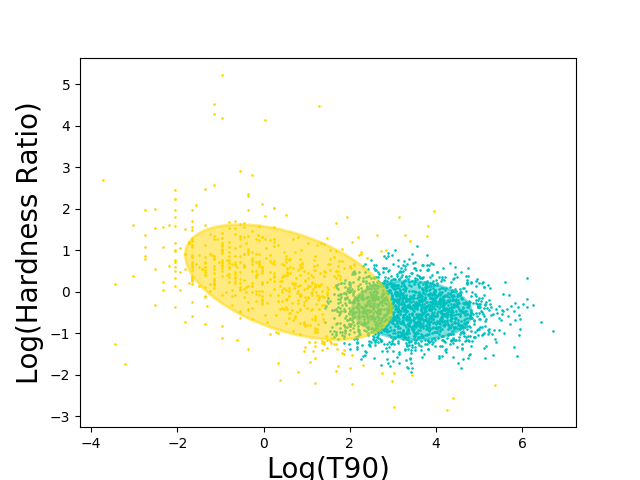}
\caption{Scatter plot of $\log (H_{obs})$ ratio vs $\log$ (T90) (expressed in seconds) for Fermi-GBM data. The ellipses indicate the 1$\sigma$ contours from our XDGMM based analysis assuming two components,  using the best-fit results tabulated in Table ~\ref{tab:fermi}.}
\label{fig:fermi1}
\end{figure}

\begin{figure}
\includegraphics[width=8cm]{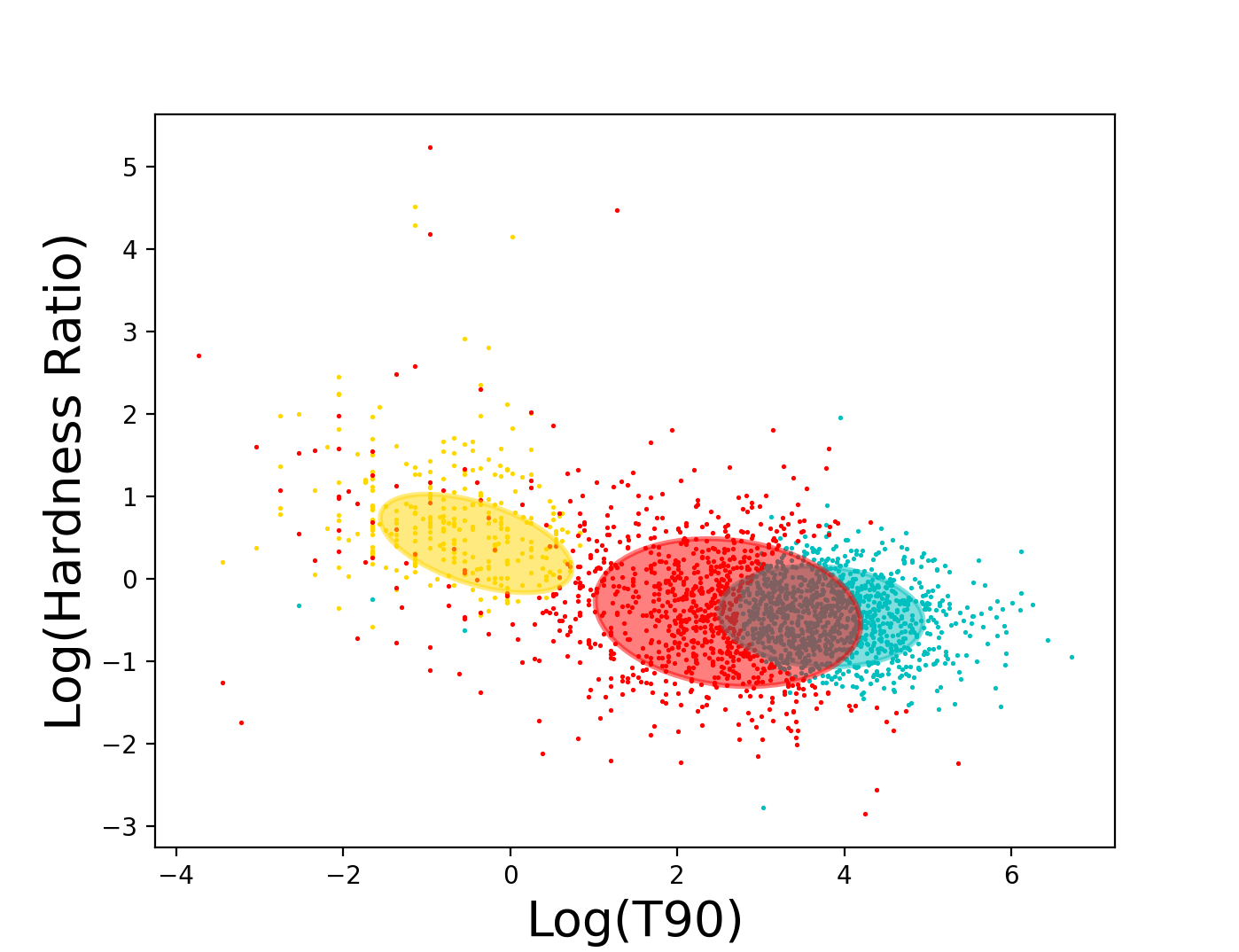}
\caption{Scatter plot of $\log H_{obs}$ ratio vs $\log$ (T90) (expressed in seconds) for Fermi-GBM data. The ellipses indicate the 1$\sigma$ contours from our XDGMM based analysis assuming three components and using  the best-fit parameters  from Table~\ref{tab:fermi}.}
\label{fig:fermi5}
\end{figure}

 A complete summary of the results on applications of XDGMM for Fermi-GBM data, including the best-fit parameters and their covariance matrices are shown in Table~\ref{tab:fermi}. For two components, we find that 360 and 1969 GRBs belong to the short  and long categories, respectively. For a three component model, we obtain  990, 268, and 1071 GRBs in short, intermediate and long GRB categories respectively.  The AIC and BIC plots as a function of the number of GRB components can be found in Fig.~\ref{fig:fermi2}.  
 The minimum value of AIC is obtained for three components, whereas for BIC the minimum value is obtained for two components. The $\Delta$ AIC values between two and three components is greater than 10, thus pointing to decisive significance in favor of three components. \rthis{We note that the AIC values for four and five components are  also smaller compared to that for two components by more than 10, which implies that four and five components are decisively favored compared  to two. However, they are greater than the AIC value for three components by more than 10, implying that AIC shows decisive significance for three components as compared to  any other number of components.}
 However, the $\Delta$BIC value between the two and three components is less than three, indicating that the difference is negligible and corresponds to ``not worth a mention'' according to Jeffreys' scale. Note however that as discussed in ~\citet{Tarnobatse19} and ~\citet{Tarnopolski19}, AIC is liberal in overfitting with a higher chance of accepting complicated models  having redundant components, than necessary. Therefore, BIC results are more trustworthy in case of a discrepancy between the two. The $1\sigma$ ellipses for the two components are shown in Fig.~\ref{fig:fermi1}. The corresponding plot for three components can be found in Fig.~\ref{fig:fermi5}.

Similar to BATSE, there is a vast amount of literature on the classification of Fermi-GBM GRBs, with no common consensus among the different works. We summarize  the results from some of these works and then compare with our results.  A summary of previous results on the classification of Fermi-GBM using durations can be found in ~\citet{Kulkarni17,Zitouni18}. All previous classification studies with Fermi GRBs show a preference for two GRBs. While this work was in progress, a short GRB detected by Fermi-GBM (GRB170817A)~\citep{GRB170817} was seen in gravitational waves (GW170817)~\citep{GW170817} creating a watershed event in the history of astronomy, and thereby opening the era of multi-messenger astronomy and providing a whole bunch of information from Astrophysics to fundamental Physics~\citep{Margutti,Woodard}.
 Subsequently, when  a clustering analysis of the GRBs from the Fermi-GBM catalog was done using multi-dimensional clustering followed by model selection using  BIC, the optimum number of clusters detected was equal to three~\citep{Horvath17}. They have also argued that  GRB170817A belongs to an intermediate class and not a short hard burst.  A follow-up cluster analysis was carried out by ~\citet{Horvathfermi2019} using 16 Fermi-GBM observables, who showed using Principal Component Analysis that the data is consistent with three GRB classes. However, the third class is different from the intermediate class. They argue that this could be caused due to the  model-dependent spectral fitting parameters provided by Fermi-GBM.
 ~\citet{Tarnopolski19} carried out   a 2-d classification using hardness and duration, and  showed that the  duration--hardness ratio plane is best represented by a mixture of two skewed \rthis{Student} $t$ distributions. ~\citet{Tarno21} showed using graph theory that the Fermi-GBM dataset is consistent with both two and three classes.
 Most recently,~\citet{Zhang22} showed using linear discriminant analysis  that a linear combination of the duration, fluence, and peak flux is a better discriminant between long and short bursts for Fermi-GBM data. They also showed that long GRBs could also be divided into long-bright GRBs and long-faint GRBs~\citep{Zhang22}.  
 
 Our analysis indicates that  AIC decisively prefers three  components, whereas for BIC the difference between the values for two and three components  is negligible.

\begin{table*}[!htbp]
\caption{Model comparison parameters for Fermi-GBM GRBs. The explanation of all the columns is same as in Table~\ref{tab:batse}. We find that both AIC  prefers three components with decisive significance, whereas BIC prefers two components albeit with marginal difference with respect to the three component model. \rthis{The AIC value for four and five components is equal to 12952 and 12954 respectively, indicating that four/five classes are favored compared to two, but disfavored when compared to three.}}
\label{tab:fermi}
\begin{tabular}{|c|ccc|cc|cc|}
\hline
$k$ & $\mu$ & $\Sigma$ & $n_{i}$ &   AIC  &  BIC   & $\Delta(AIC) $& $ \Delta(BIC)$ \\
\hline
\multirow{2}{*}{2} & (3.27,-0.46) &  $\left(\begin{array}{cc}                            1.11 &-0.04\\-0.04 &0.28 \\               \end{array}\right)$ & 1969       & \multirow{2}{*}{12973.5} & \multirow{2}{*}{\textbf{13036.8}} & \multirow{5}{*}{32.3} & \multirow{5}{*}{-2.2} \\
\cline{2-4}
& (-0.09,0.34) &$\left(\begin{array}{cc}       0.97 &-0.19\\-0.19&0.23 \\     \end{array}\right)$ & 360 & &  & &\\
\cline{1-6}
\multirow{3}{*}{3} &(3.71,-0.46) &$\left(\begin{array}{cc}       0.6334 &-0.1730\\ -0.1730 &0.1744 \\     \end{array}\right)$ & 1071    & \multirow{3}{*}{\textbf{12941}} & \multirow{3}{*}{13039}  & & \\
\cline{2-4}
& (-0.43,0.44) &$\left(\begin{array}{cc}       0.6334 &-0.1730\\-0.1730& 0.1744 \\     \end{array}\right)$ & 990 & &  & & \\
\cline{2-4}
& (2.59,-0.41) &$\left(\begin{array}{cc}       1.2993&-0.1167\\ -0.1167&0.4048 \\     \end{array}\right)$&268 & &  & & \\
\hline
\end{tabular}
%\tablecomments{In the table the preferred value for every test is highlighted in bold.}
\end{table*}

\iffalse

\fi

\begin{table*}[t]
\caption{Summary of model  comparison tests for  the  GRB datasets analyzed.  For each dataset we have indicated the preferred model based on  the difference between the values of the AIC and BIC between the best component and the one with the higher value.}
\label{tab:summ}
\begin{tabular}{|c|cc|cc|}
\hline
\multirow{2}{*}{Dataset}  & \multicolumn{2}{c}{$\Delta AIC$\
} & \multicolumn{2}{|c|}{$\Delta BIC$} \\
\cline{2-5}
&  Optimum Classes  & Difference & Optimum Classes   & Difference \\
\hline
BATSE & 2 & 9.5 & 2 & 42.9 \\
\hline
Fermi & 3 & 32.3 & 2 & 2.2 \\
\hline
%Swift & 3 & 22 & 2 & 8 \\
%\hline
%Intrinsic Swift  & 3 & 19 & 2 & 4.1 \\
%\hline
\end{tabular}
\end{table*}

\section{Conclusions}
\label{sec:conclusions}
The main goal of this work was to find the optimum number of  GRB components   by carrying out a  two-dimensional clustering in the T90 vs hardness plane, along with 
 incorporating the errors in T90 and hardness in the aforementioned analysis. Although there are a plethora of works over a time span of more than two decades, which have done a multidimensional classification using multiple GRB observables, none of them have incorporated the  uncertainties in the analysis. This is the first work on GRB classification which has included the aforementioned uncertainties. For our analysis, we used  an extension of the GMM based classification which incorporates the uncertainties,  referred to as XDGMM~\citep{Bovy}. We used the data from two space-based detectors, BATSE and  Fermi-GBM for our analysis.

We then used two information criterion  based statistical tests to ascertain the optimum number of GRB classes in both the datasets.  These tests include AIC and BIC model comparison tests. The statistical significance from the information criterion based tests was obtained qualitatively using empirical strength of evidence rules~\citep{Shi,Krishak}.   The AIC/BIC trends as a function of number of components for BATSE and Fermi-GBM can be found in Fig.~\ref{fig:batse2} and Fig.~\ref{fig:fermi2}, respectively. The best-fit XDGMM values for BATSE and Fermi-GBM are summarized in Table~\ref{tab:batse} and Table~\ref{tab:fermi}, respectively. A tabular summary of our results on optimum number of components can be found in Table~\ref{tab:summ}.  Our main conclusions for both the datasets are as follows:

\begin{itemize}
\item For BATSE, we find that both AIC and BIC prefer two components with very strong (AIC) or decisive significance (BIC).

\item For Fermi-GBM,  AIC prefers three components with decisive significance. However, BIC prefers two components, albeit with very marginal significance. Since AIC is known to be liberal in overfitting more complicated models, the results from BIC should be trusted in this case~\citep{Tarnopolski19}.

\end{itemize}

We should note  that ever since it was shown that $\log$ (T90) can be adequately modelled by sum of normal distributions~\citep{McBreen,Koshut}, almost all literature on GRB classification has assumed that lognormal distributions  can adequately describe the distribution of various GRB observables, and this work is no exception. However, if the distribution of $\log$ (T90) is not comprised of normal distributions and is  skew-symmetric~\citep{Koen,Tarnopolski15,Tarnopolski19}, then one would also need to replace the  Gaussian  distributions  with skew-symmetric distributions, in addition to incorporating the uncertainties. We defer such an analysis to a future work.

%\item When we looked at Swift GRBs in the observer frame, we find that AIC and BIC reach opposite conclusions. AIC prefers three components with decisive significance. BIC shows a prefer for two components, albeit with strong significance. The minimum value of AIC is obtained for five components.

%\item When we repeat the analysis for SwiftGRBs in the intrinsic frame, we reach the same conclusions as in the observer frame. AIC prefers three components with decisive significance, where BIC prefers two with strong evidence.

We note that all our codes to reproduce these results have been uploaded on the web at  \url{https://github.com/lostsoul3/GRB_analysis}

\section{Acknowledgements}
Aishwarya Bhave was supported by Microsoft Internship program at IIT Hyderabad.
We are grateful to P. Narayana Bhat for providing us the  hardness data for Fermi-GBM GRBs and also the anonymous referee for constructive feedback on the manuscript.

\iffalse
\section*{Declaration}
\begin{itemize}
    \item \textbf{Funding:} AB was supported by Microsoft internship program during Winter of 2016.
    \item \textbf{Competing Interests:} The authors have no relevant financial or non-financial interests to disclose.
    \item \textbf{Author Contributions:} All authors contributed to the material presented in this paper. All authors read and approved the final manuscript.
    \item \textbf{Data availability:} The datasets generated during and/or analysed during the current study have been uploaded on github and  are  also available from the corresponding author on reasonable request.
\end{itemize}
\fi
%\bibliographystyle{spr-mp-nameyear-cnd}
\bibliographystyle{apj}
\bibliography{grb2d}

\begin{thebibliography}{}
\expandafter\ifx\csname natexlab\endcsname\relax\def\natexlab#1{#1}\fi

\bibitem[{{Abbott} {et~al.}(2017){Abbott}, {Abbott}, {Abbott}, {Acernese},
  {Ackley}, {Adams}, {Adams}, {Addesso}, {Adhikari}, {Adya}, \&
  et~al.}]{GW170817}
{Abbott}, B.~P., {Abbott}, R., {Abbott}, T.~D., {et~al.} 2017, Physical Review
  Letters, 119, 161101

\bibitem[{{Ahumada} {et~al.}(2021){Ahumada}, {Singer}, {Anand}, {Coughlin},
  {Kasliwal}, {Ryan}, {Andreoni}, {Cenko}, {Fremling}, {Kumar}, {Pang},
  {Burns}, {Cunningham}, {Dichiara}, {Dietrich}, {Svinkin}, {Almualla},
  {Castro-Tirado}, {De}, {Dunwoody}, {Gatkine}, {Hammerstein}, {Iyyani},
  {Mangan}, {Perley}, {Purkayastha}, {Bellm}, {Bhalerao}, {Bolin}, {Bulla},
  {Cannella}, {Chandra}, {Duev}, {Frederiks}, {Gal-Yam}, {Graham}, {Ho},
  {Hurley}, {Karambelkar}, {Kool}, {Kulkarni}, {Mahabal}, {Masci}, {McBreen},
  {Pandey}, {Reusch}, {Ridnaia}, {Rosnet}, {Rusholme}, {Carracedo}, {Smith},
  {Soumagnac}, {Stein}, {Troja}, {Tsvetkova}, {Walters}, \& {Valeev}}]{Ahumada}
{Ahumada}, T., {Singer}, L.~P., {Anand}, S., {et~al.} 2021, Nature Astronomy,
  5, 917

\bibitem[{{Amati}(2021)}]{Amati20}
{Amati}, L. 2021, Nature Astronomy, 5, 877

\bibitem[{{Band}(2006)}]{Band06}
{Band}, D.~L. 2006, \apj, 644, 378

\bibitem[{{Boran} {et~al.}(2018){Boran}, {Desai}, {Kahya}, \&
  {Woodard}}]{Woodard}
{Boran}, S., {Desai}, S., {Kahya}, E.~O., \& {Woodard}, R.~P. 2018, \prd, 97,
  041501

\bibitem[{{Bovy} {et~al.}(2011){Bovy}, {Hogg}, \& {Roweis}}]{Bovy}
{Bovy}, J., {Hogg}, D.~W., \& {Roweis}, S.~T. 2011, Annals of Applied
  Statistics, 5, 1657

\bibitem[{{Bromberg} {et~al.}(2013){Bromberg}, {Nakar}, {Piran}, \&
  {Sari}}]{Bromberg}
{Bromberg}, O., {Nakar}, E., {Piran}, T., \& {Sari}, R. 2013, \apj, 764, 179

\bibitem[{Burnham \& Anderson(2004)}]{Burnham}
Burnham, K.~P., \& Anderson, D.~R. 2004, Sociological methods \& research, 33,
  261

\bibitem[{{Chattopadhyay} \& {Maitra}(2017)}]{Chattopadhyay17}
{Chattopadhyay}, S., \& {Maitra}, R. 2017, \mnras, 469, 3374

\bibitem[{{Chattopadhyay} \& {Maitra}(2018)}]{Maitra17}
---. 2018, \mnras, 481, 3196

\bibitem[{{Chattopadhyay} {et~al.}(2007{\natexlab{a}}){Chattopadhyay}, {Misra},
  {Chattopadhyay}, \& {Naskar}}]{Chattopadhyay}
{Chattopadhyay}, T., {Misra}, R., {Chattopadhyay}, A.~K., \& {Naskar}, M.
  2007{\natexlab{a}}, \apj, 667, 1017

\bibitem[{{Chattopadhyay} {et~al.}(2007{\natexlab{b}}){Chattopadhyay}, {Misra},
  {Chattopadhyay}, \& {Naskar}}]{Misra}
---. 2007{\natexlab{b}}, \apj, 667, 1017

\bibitem[{{Coronado-Bl{\'a}zquez} {et~al.}(2019){Coronado-Bl{\'a}zquez},
  {S{\'a}nchez-Conde}, {Di Mauro}, {Aguirre-Santaella}, {Ciuc{\u{a}}},
  {Dom{\'\i}nguez}, {Kawata}, \& {Mirabal}}]{Miguel}
{Coronado-Bl{\'a}zquez}, J., {S{\'a}nchez-Conde}, M.~A., {Di Mauro}, M.,
  {et~al.} 2019, \jcap, 2019, 045

\bibitem[{{Coward} {et~al.}(2013){Coward}, {Howell}, {Branchesi}, {Stratta},
  {Guetta}, {Gendre}, \& {Macpherson}}]{Coward}
{Coward}, D.~M., {Howell}, E.~J., {Branchesi}, M., {et~al.} 2013, \mnras, 432,
  2141

\bibitem[{{Desai}(2016)}]{Desai16b}
{Desai}, S. 2016, EPL (Europhysics Letters), 115, 20006

\bibitem[{{Desai} \& {Liu}(2016)}]{Desai16a}
{Desai}, S., \& {Liu}, D.~W. 2016, Astroparticle Physics, 82, 86

\bibitem[{Ganguly \& Desai(2017)}]{Ganguly17}
Ganguly, S., \& Desai, S. 2017, Astroparticle Physics, C94, 17

\bibitem[{Ghosh \& Sen(1984)}]{Ghosh85}
Ghosh, J.~K., \& Sen, P.~K. 1984, On the asymptotic performance of the log
  likelihood ratio statistic for the mixture model and related results
  (University of North Carolina at Chapel Hill. Institute of Statistics)

\bibitem[{{Goldstein} {et~al.}(2017){Goldstein}, {Veres}, {Burns}, {Briggs},
  {Hamburg}, {Kocevski}, {Wilson-Hodge}, {Preece}, {Poolakkil}, {Roberts},
  {Hui}, {Connaughton}, {Racusin}, {von Kienlin}, {Dal Canton}, {Christensen},
  {Littenberg}, {Siellez}, {Blackburn}, {Broida}, {Bissaldi}, {Cleveland},
  {Gibby}, {Giles}, {Kippen}, {McBreen}, {McEnery}, {Meegan}, {Paciesas}, \&
  {Stanbro}}]{GRB170817}
{Goldstein}, A., {Veres}, P., {Burns}, E., {et~al.} 2017, \apjl, 848, L14

\bibitem[{{Gruber} {et~al.}(2014){Gruber}, {Goldstein}, {Weller von Ahlefeld},
  {Narayana Bhat}, {Bissaldi}, {Briggs}, {Byrne}, {Cleveland}, {Connaughton},
  {Diehl}, {Fishman}, {Fitzpatrick}, {Foley}, {Gibby}, {Giles}, {Greiner},
  {Guiriec}, {van der Horst}, {von Kienlin}, {Kouveliotou}, {Layden}, {Lin},
  {Meegan}, {McGlynn}, {Paciesas}, {Pelassa}, {Preece}, {Rau}, {Wilson-Hodge},
  {Xiong}, {Younes}, \& {Yu}}]{Gruber}
{Gruber}, D., {Goldstein}, A., {Weller von Ahlefeld}, V., {et~al.} 2014, \apjs,
  211, 12

\bibitem[{{Holoien} {et~al.}(2017){Holoien}, {Marshall}, \&
  {Wechsler}}]{Wechsler}
{Holoien}, T. W.~S., {Marshall}, P.~J., \& {Wechsler}, R.~H. 2017, \aj, 153,
  249

\bibitem[{{Horv{\'a}th}(1998)}]{Horvath98}
{Horv{\'a}th}, I. 1998, \apj, 508, 757

\bibitem[{{Horv{\'a}th}(2002)}]{Horvath2002}
---. 2002, \aap, 392, 791

\bibitem[{{Horv{\'a}th} {et~al.}(2010){Horv{\'a}th}, {Bagoly}, {Bal{\'a}zs},
  {de Ugarte Postigo}, {Veres}, \& {M{\'e}sz{\'a}ros}}]{Horvath10}
{Horv{\'a}th}, I., {Bagoly}, Z., {Bal{\'a}zs}, L.~G., {et~al.} 2010, \apj, 713,
  552

\bibitem[{{Horv{\'a}th} {et~al.}(2006){Horv{\'a}th}, {Bal{\'a}zs}, {Bagoly},
  {Ryde}, \& {M{\'e}sz{\'a}ros}}]{Horvath06}
{Horv{\'a}th}, I., {Bal{\'a}zs}, L.~G., {Bagoly}, Z., {Ryde}, F., \&
  {M{\'e}sz{\'a}ros}, A. 2006, \aap, 447, 23

\bibitem[{{Horv{\'a}th} {et~al.}(2008){Horv{\'a}th}, {Bal{\'a}zs}, {Bagoly}, \&
  {Veres}}]{Horvath2008}
{Horv{\'a}th}, I., {Bal{\'a}zs}, L.~G., {Bagoly}, Z., \& {Veres}, P. 2008,
  \aap, 489, L1

\bibitem[{{Horv{\'a}th} {et~al.}(2019){Horv{\'a}th}, {Hakkila}, {Bagoly},
  {T{\'o}th}, {R{\'a}cz}, {Pint{\'e}r}, \& {T{\'o}th}}]{Horvathfermi2019}
{Horv{\'a}th}, I., {Hakkila}, J., {Bagoly}, Z., {et~al.} 2019, \apss, 364, 105

\bibitem[{{Horv{\'a}th} \& {T{\'o}th}(2016)}]{Horvath16}
{Horv{\'a}th}, I., \& {T{\'o}th}, B.~G. 2016, \apss, 361, 155

\bibitem[{{Horv{\'a}th} {et~al.}(2018){Horv{\'a}th}, {T{\'o}th}, {Hakkila},
  {T{\'o}th}, {Bal{\'a}zs}, {R{\'a}cz}, {Pint{\'e}r}, \& {Bagoly}}]{Horvath17}
{Horv{\'a}th}, I., {T{\'o}th}, B.~G., {Hakkila}, J., {et~al.} 2018, \apss, 363,
  53

\bibitem[{{Huja} {et~al.}(2009){Huja}, {M{\'e}sz{\'a}ros}, \& {{\v
  R}{\'{\i}}pa}}]{Huja09}
{Huja}, D., {M{\'e}sz{\'a}ros}, A., \& {{\v R}{\'{\i}}pa}, J. 2009, \aap, 504,
  67

\bibitem[{{Ivezi{\'c}} {et~al.}(2014){Ivezi{\'c}}, {Connolly}, {Vanderplas}, \&
  {Gray}}]{astroML}
{Ivezi{\'c}}, {\v Z}., {Connolly}, A., {Vanderplas}, J., \& {Gray}, A. 2014,
  Statistics, Data Mining and Machine Learning in Astronomy (Princeton
  University Press)

\bibitem[{{Jespersen} {et~al.}(2020){Jespersen}, {Severin}, {Steinhardt},
  {Vinther}, {Fynbo}, {Selsing}, \& {Watson}}]{Jespersen}
{Jespersen}, C.~K., {Severin}, J.~B., {Steinhardt}, C.~L., {et~al.} 2020,
  \apjl, 896, L20

\bibitem[{Kass \& Raftery(1995)}]{Robert}
Kass, R.~E., \& Raftery, A.~E. 1995, Journal of the American Statistical
  Association, 90, 773

\bibitem[{{Keitel}(2019)}]{Keitel}
{Keitel}, D. 2019, \mnras, 485, 1665

\bibitem[{{Kerscher} \& {Weller}(2019)}]{Weller}
{Kerscher}, M., \& {Weller}, J. 2019, SciPost Physics Lecture Notes, 9,
  arXiv:1901.07726

\bibitem[{{Koen} \& {Bere}(2012)}]{Koen}
{Koen}, C., \& {Bere}, A. 2012, \mnras, 420, 405

\bibitem[{{Koposov} {et~al.}(2013){Koposov}, {Belokurov}, \&
  {Evans}}]{sagittarius_koposov}
{Koposov}, S.~E., {Belokurov}, V., \& {Evans}, N.~W. 2013, \apj, 766, 79

\bibitem[{{Koshut} {et~al.}(1996){Koshut}, {Paciesas}, {Kouveliotou}, {van
  Paradijs}, {Pendleton}, {Fishman}, \& {Meegan}}]{Koshut}
{Koshut}, T.~M., {Paciesas}, W.~S., {Kouveliotou}, C., {et~al.} 1996, \apj,
  463, 570

\bibitem[{{Kouveliotou} {et~al.}(1993){Kouveliotou}, {Meegan}, {Fishman},
  {Bhat}, {Briggs}, {Koshut}, {Paciesas}, \& {Pendleton}}]{Kouveliotou}
{Kouveliotou}, C., {Meegan}, C.~A., {Fishman}, G.~J., {et~al.} 1993, \apjl,
  413, L101

\bibitem[{{Krishak} {et~al.}(2020){Krishak}, {Dantuluri}, \&
  {Desai}}]{KrishakDAMA}
{Krishak}, A., {Dantuluri}, A., \& {Desai}, S. 2020, \jcap, 2020, 007

\bibitem[{{Krishak} \& {Desai}(2019)}]{KrishakOJA}
{Krishak}, A., \& {Desai}, S. 2019, The Open Journal of Astrophysics, 2, E12

\bibitem[{{Krishak} \& {Desai}(2020{\natexlab{a}})}]{Krishakanais}
---. 2020{\natexlab{a}}, Progress of Theoretical and Experimental Physics,
  2020, 093F01

\bibitem[{{Krishak} \& {Desai}(2020{\natexlab{b}})}]{Krishak}
---. 2020{\natexlab{b}}, \jcap, 2020, 006

\bibitem[{{Kuhn} \& {Feigelson}(2017)}]{Feigelson17}
{Kuhn}, M.~A., \& {Feigelson}, E.~D. 2017, ArXiv e-prints, arXiv:1711.11101

\bibitem[{{Kulkarni} \& {Desai}(2017)}]{Kulkarni17}
{Kulkarni}, S., \& {Desai}, S. 2017, \apss, 362, 70

\bibitem[{{Kulkarni} \& {Desai}(2018)}]{Kulkarniexoplanet}
---. 2018, The Open Journal of Astrophysics, 1, 4

\bibitem[{{Kumar} \& {Zhang}(2015)}]{Kumar15}
{Kumar}, P., \& {Zhang}, B. 2015, \physrep, 561, 1

\bibitem[{{Kwong} \& {Nadarajah}(2018)}]{Kwong}
{Kwong}, H.~S., \& {Nadarajah}, S. 2018, \mnras, 473, 625

\bibitem[{{Liddle}(2004)}]{Liddle}
{Liddle}, A.~R. 2004, \mnras, 351, L49

\bibitem[{{Liddle}(2007)}]{Liddle07}
---. 2007, \mnras, 377, L74

\bibitem[{{Lyons}(2016)}]{Lyons}
{Lyons}, L. 2016, ArXiv e-prints, arXiv:1607.03549

\bibitem[{{Margutti} \& {Chornock}(2021)}]{Margutti}
{Margutti}, R., \& {Chornock}, R. 2021, \araa, 59, arXiv:2012.04810

\bibitem[{{McBreen} {et~al.}(1994){McBreen}, {Hurley}, {Long}, \&
  {Metcalfe}}]{McBreen}
{McBreen}, B., {Hurley}, K.~J., {Long}, R., \& {Metcalfe}, L. 1994, \mnras,
  271, 662

\bibitem[{{Mukherjee} {et~al.}(1998){Mukherjee}, {Feigelson}, {Jogesh Babu},
  {Murtagh}, {Fraley}, \& {Raftery}}]{Mukherjee}
{Mukherjee}, S., {Feigelson}, E.~D., {Jogesh Babu}, G., {et~al.} 1998, \apj,
  508, 314

\bibitem[{{Nakar}(2007)}]{Nakar}
{Nakar}, E. 2007, \physrep, 442, 166

\bibitem[{{Narayana Bhat} {et~al.}(2016){Narayana Bhat}, {Meegan}, {von
  Kienlin}, {Paciesas}, {Briggs}, {Burgess}, {Burns}, {Chaplin}, {Cleveland},
  {Collazzi}, {Connaughton}, {Diekmann}, {Fitzpatrick}, {Gibby}, {Giles},
  {Goldstein}, {Greiner}, {Jenke}, {Kippen}, {Kouveliotou}, {Mailyan},
  {McBreen}, {Pelassa}, {Preece}, {Roberts}, {Sparke}, {Stanbro}, {Veres},
  {Wilson-Hodge}, {Xiong}, {Younes}, {Yu}, \& {Zhang}}]{GBM}
{Narayana Bhat}, P., {Meegan}, C.~A., {von Kienlin}, A., {et~al.} 2016, \apjs,
  223, 28

\bibitem[{{Paciesas} {et~al.}(1999){Paciesas}, {Meegan}, {Pendleton}, {Briggs},
  {Kouveliotou}, {Koshut}, {Lestrade}, {McCollough}, {Brainerd}, {Hakkila},
  {Henze}, {Preece}, {Connaughton}, {Kippen}, {Mallozzi}, {Fishman},
  {Richardson}, \& {Sahi}}]{BATSE}
{Paciesas}, W.~S., {Meegan}, C.~A., {Pendleton}, G.~N., {et~al.} 1999, \apjs,
  122, 465

\bibitem[{{Perna} {et~al.}(2018){Perna}, {Lazzati}, \& {Cantiello}}]{Perna18}
{Perna}, R., {Lazzati}, D., \& {Cantiello}, M. 2018, \apj, 859, 48

\bibitem[{{Reddy Ch.} \& {Desai}(2022)}]{Reddy}
{Reddy Ch.}, T.~T., \& {Desai}, S. 2022, \na, 91, 101673

\bibitem[{{Schady}(2017)}]{Schady}
{Schady}, P. 2017, Royal Society Open Science, 4, 170304

\bibitem[{{Sharma}(2017)}]{Sharma}
{Sharma}, S. 2017, \araa, 55, 213

\bibitem[{{Shi} {et~al.}(2012){Shi}, {Huang}, \& {Lu}}]{Shi}
{Shi}, K., {Huang}, Y.~F., \& {Lu}, T. 2012, \mnras, 426, 2452

\bibitem[{{Singh} \& {Desai}(2022)}]{Singh}
{Singh}, A., \& {Desai}, S. 2022, \jcap, 2022, 010

\bibitem[{{Tarnopolski}(2015{\natexlab{a}})}]{Tarnopolski15}
{Tarnopolski}, M. 2015{\natexlab{a}}, \aap, 581, A29

\bibitem[{{Tarnopolski}(2015{\natexlab{b}})}]{Tatro15}
---. 2015{\natexlab{b}}, \apss, 359, 20

\bibitem[{{Tarnopolski}(2016{\natexlab{a}})}]{Tarnopolski16b}
---. 2016{\natexlab{a}}, \apss, 361, 125

\bibitem[{{Tarnopolski}(2016{\natexlab{b}})}]{Tarnopolski}
---. 2016{\natexlab{b}}, \na, 46, 54

\bibitem[{{Tarnopolski}(2019{\natexlab{a}})}]{Tarnosai}
---. 2019{\natexlab{a}}, \memsai, 90, 45

\bibitem[{{Tarnopolski}(2019{\natexlab{b}})}]{Tarnopolski19}
---. 2019{\natexlab{b}}, \apj, 870, 105

\bibitem[{{Tarnopolski}(2019{\natexlab{c}})}]{Tarnobatse19}
---. 2019{\natexlab{c}}, \apj, 887, 97

\bibitem[{{Tarnopolski}(2022)}]{Tarno21}
---. 2022, \aap, 657, A13

\bibitem[{{T{\'o}th} {et~al.}(2019){T{\'o}th}, {R{\'a}cz}, \&
  {Horv{\'a}th}}]{Horvath19}
{T{\'o}th}, B.~G., {R{\'a}cz}, I.~I., \& {Horv{\'a}th}, I. 2019, \mnras, 486,
  4823

\bibitem[{{Veres} {et~al.}(2010){Veres}, {Bagoly}, {Horv{\'a}th}, {M{\'e}sz{\\
  'a}ros}, \& {Bal{\'a}zs}}]{Veres}
{Veres}, P., {Bagoly}, Z., {Horv{\'a}th}, I., {M{\'e}sz{\\ 'a}ros}, A., \&
  {Bal{\'a}zs}, L.~G. 2010, \apj, 725, 1955

\bibitem[{{von Kienlin} {et~al.}(2014){von Kienlin}, {Meegan}, {Paciesas},
  {Bhat}, {Bissaldi}, {Briggs}, {Burgess}, {Byrne}, {Chaplin}, {Cleveland},
  {Connaughton}, {Collazzi}, {Fitzpatrick}, {Foley}, {Gibby}, {Giles},
  {Goldstein}, {Greiner}, {Gruber}, {Guiriec}, {van der Horst}, {Kouveliotou},
  {Layden}, {McBreen}, {McGlynn}, {Pelassa}, {Preece}, {Rau}, {Tierney},
  {Wilson-Hodge}, {Xiong}, {Younes}, \& {Yu}}]{VonKienlin}
{von Kienlin}, A., {Meegan}, C.~A., {Paciesas}, W.~S., {et~al.} 2014, \apjs,
  211, 13

\bibitem[{{von Kienlin} {et~al.}(2020){von Kienlin}, {Meegan}, {Paciesas},
  {Bhat}, {Bissaldi}, {Briggs}, {Burns}, {Cleveland}, {Gibby}, {Giles},
  {Goldstein}, {Hamburg}, {Hui}, {Kocevski}, {Mailyan}, {Malacaria},
  {Poolakkil}, {Preece}, {Roberts}, {Veres}, \& {Wilson-Hodge}}]{fermi2020}
---. 2020, \apj, 893, 46

\bibitem[{{Woosley} \& {Bloom}(2006)}]{Woosley}
{Woosley}, S.~E., \& {Bloom}, J.~S. 2006, \araa, 44, 507

\bibitem[{{Yang} {et~al.}(2016){Yang}, {Zhang}, \& {Jiang}}]{Jiang}
{Yang}, E.~B., {Zhang}, Z.~B., \& {Jiang}, X.~X. 2016, \apss, 361, 257

\bibitem[{{Zhang} {et~al.}(2016{\natexlab{a}}){Zhang}, {L{\"u}}, \&
  {Liang}}]{Zhang16}
{Zhang}, B., {L{\"u}}, H.-J., \& {Liang}, E.-W. 2016{\natexlab{a}}, \ssr, 202,
  3

\bibitem[{{Zhang} {et~al.}(2009){Zhang}, {Zhang}, {Virgili}, {Liang}, {Kann},
  {Wu}, {Proga}, {Lv}, {Toma}, {M{\'e}sz{\'a}ros}, {Burrows}, {Roming}, \&
  {Gehrels}}]{Zhang}
{Zhang}, B., {Zhang}, B.-B., {Virgili}, F.~J., {et~al.} 2009, \apj, 703, 1696

\bibitem[{{Zhang} {et~al.}(2022){Zhang}, {Shao}, {Zhang}, {Zou}, {Sun}, {Yao},
  \& {Li}}]{Zhang22}
{Zhang}, S., {Shao}, L., {Zhang}, B.-B., {et~al.} 2022, \apj, 926, 170

\bibitem[{{Zhang} \& {Choi}(2008)}]{Zhang08}
{Zhang}, Z.-B., \& {Choi}, C.-S. 2008, \aap, 484, 293

\bibitem[{{Zhang} {et~al.}(2016{\natexlab{b}}){Zhang}, {Yang}, {Choi}, \&
  {Chang}}]{Yang}
{Zhang}, Z.-B., {Yang}, E.-B., {Choi}, C.-S., \& {Chang}, H.-Y.
  2016{\natexlab{b}}, \mnras, 462, 3243

\bibitem[{{Zitouni} {et~al.}(2018){Zitouni}, {Guessoum}, {AlQassimi}, \&
  {Alaryani}}]{Zitouni18}
{Zitouni}, H., {Guessoum}, N., {AlQassimi}, K.~M., \& {Alaryani}, O. 2018,
  \apss, 363, 223

\bibitem[{{Zitouni} {et~al.}(2015){Zitouni}, {Guessoum}, {Azzam}, \&
  {Mochkovitch}}]{Zitouni}
{Zitouni}, H., {Guessoum}, N., {Azzam}, W.~J., \& {Mochkovitch}, R. 2015,
  \apss, 357, 7

\end{thebibliography}
%\bibliography{biblio-u1}

\end{document}